\documentclass [a4paper]{article}
\usepackage {isolatin1,latexsym,a4}

\begin {document}

\sloppy

\hoffset -1cm \voffset -1cm 

\pagestyle {myheadings}
\markright{\large Matrix Representation of Special Relativity \hfill \small \today \hspace*{1cm}}

\newcommand{\FN}[1]{\footnote { #1} } 
\newcommand{\MBOX}[1]{\quad \mbox {#1} \quad}
\newcommand{\CC}[1]{#1^\ast} 
\newcommand{\HC}[1]{#1^\dagger} 
\newcommand{\HCB}[1]{\widetilde{#1}}
\newcommand{\MV}[1]{{\bf #1}} 
\newcommand{\TR}[1] {\mathcal T( #1 )} 
\newcommand{\LAGR} {\mathcal L} 
\newcommand{\BE}{\begin{equation}}
\newcommand{\EE}{\end{equation}}
\newcommand{\SC}[1]{{\sc #1}} 

\newcommand{\PD}[2][]{\frac{\partial #1}{\partial #2}} 
\newcommand{\DEF} {\stackrel {def} = }

\title{ Matrix Representation of Special Relativity }

\author{Wolfgang Köhler, Potsdam, Germany\FN{E-mail address: {\tt wolfk@gfz-potsdam.de} 
    \hfill Homepage: {\tt http://icgem.gfz-potsdam.de/QM} }}

\maketitle

\begin{abstract}
   I compare the matrix representation of the basic statements of Special Relativity 
   with the conventional vector space representation.\\
   It is shown, that the matrix form reproduces all equations in a very concise and elegant form, 
   namely:   \SC{Maxwell} equations, \SC{Lorentz}-force, energy-momentum tensor, 
   \SC{Dirac}-equation and \SC{Lagrangian}s.\\
   The matrix representation requires
   fewer assumptions, uses fewer parameters and may lead to new insights into physical reality.\\
   A new result is a matrix form of \SC{Dirac}'s equation. 
   It can explain the non-existence of right-handed neutrinos and 
   can be generalized to include a new variant of 
   \SC{Yang-Mills} gauge fields, which possibly express unified electro-weak interactions.
\end{abstract}
\tableofcontents

\section{Introduction}

The possibility of representing \SC{Minkowski} spacetime vectors with 2x2-matrices has been known 
since the 1920ies (e.g. \cite{VDW}, \cite{LL}, pp. 61). 
It is a consequence of the fact, that the \SC{Lorentz}-group is homomorphic to the group of
unimodular binary matrices $SL(2,\mathcal C)$.

This matrix representation is mostly used to show, how covariant equations for spinors
can be derived.
There is a general consensus, that both representations
(matrix form and usual component form) are actually equivalent methods to express the
equations of Special Relativity, and the matrix form is used very rarely in publications.\FN{
  One of the first fundamental papers on this topic is \cite{WEYL}, 
  where the idea of {\em em. gauge symmetry} was invented and some of the concepts and eqs. 
  below can be found. 
  However, he focuses there on gravitation and curved space-time.\\
  A newer perspective can be found in \cite{PENROSE}, where the authors try to 
  give a fundamental overview from a mathematical point of view. They also focus
  on GR and various kinds of generalizations.\\
  In neither of both works the rigorous physical interpretation is adopted, 
  which is proposed here.
}

One principal reason for this is, that conventional component formulae can be
formally applied to an arbitrary number of dimensions, 
while the matrix form is only possible for the {\em four-dimensional} case. 

In this article I show, that the matrix form gives an elegant means to derive all
equations of special relativity but with significantly less prerequisites.

The most important prerequsite is the existence of a {\em metric tensor} with 
the signature $(+ - -\; -)$, that has to be postulated for the vector space 
(in principle, any metric signature would be conceivable). 
This metric is automatically determined for the matrix formalism. 

To value this fact, one should note, that the metric tensor is - at least implicitly -
contained in every relativistic equation. 

Moreover, the homogeneous \SC{Maxwell} equations, which have to be  introduced independently in SRT in 
component form, are a direct consequence of the inhomogeneous \SC{Maxwell} eqs. here.

The last and most important argument gives the reformulation of the \SC{Dirac} equation in matrix form.
All arbitrary free parameters without physical content, which arise in the 4-spinor form, vanish here,
because the remaining similarity transformations can be understood as {\em gauge transformations}.\\

Thus, the main aim of this paper is a {\em change of perspective}: 
Physical spacetime is primarily to represent by a {\em matrix algebra} and the component 
formulation is a derived one, which also has its disadvantages.\\
I will denote this perspective as "matrix spacetime" (MST) compared to "vector spacetime".\FN{
  Compare e.g. \cite{HESTENES}, where a similar concept with a four-dimensional algebra
  based on the \SC{Clifford}-matrices is presented. He uses a similar term "spacetime algebra" 
  (STA).
}

Please note that, if this point of view is adopted, this is {\em not only a formal aspect}, 
but it has far-reaching consequences for many other physical theories.
E.g. obviously all theories with more than four spacetime dimensions are excluded.
The matrix form also has applicability in General Relativity, but this goes beyond the scope of this paper.\FN{
  Then the 4 basis matrices \{$\tau_\mu$\} introduced below, or equivalently the 16 coefficients $a_\mu^\nu$, 
  which play the role of {\em tetrades}, have to be used instead the metric tensor as field variables. 
  More detailled discussions of this can be 
  found again in \cite{WEYL} and \cite{PENROSE}
}\\


This new perspective also may lead to new theories, e.g. if possible generalizations of this 
form are considered. One might also look for an underlying {\em spinor structure} for the matrix algebra, 
which is e.g. the main thesis of the ``twistor-theory'' presented in \cite{PENROSE} (Vol. II) 
but has not led to a satisfactory physical theory yet.\\

In conclusion I have to say, that many of the equations presented here, can also be found scattered 
in other publications. 

New in any case, is the notation of \SC{Dirac}s eq. as ``matrix equation''. 
Also the corresponding \SC{Lagrangian}, I have not found in another publication.
This new form perhaps allows new insights in particle physics, esp. {\em unified ectro-weak theory}.\\


\section{Matrix Representation of \SC{Minkowski}-Vectors}
Let me start with the 4-dimensional vector space of real numbers $V^4 = \{(x^0,x^1,x^2,x^3)\}$.
This can be mapped one-to-one to the set of hermitean 2x2 matrices $\MV M =\{\MV x \}$, 
when a basis of 4 linearly independent hermitean matrices $\tau_\mu = (\tau_0,\dots \tau_3)$ is given by
(as usual, over double upper and lower indices $\mu = 0,\dots 3$ is to sum):
\begin{equation}
  \label{eq-map}
  \MV x =  x^\mu \tau_\mu.
\end{equation}
These hermitean matrices $\MV x$ build a well defined subset of the binary matrix algebra.
In the following they  are denoted as {\em \SC{Minkowski}-matrices} and represented by {\bf boldface}
letters (except the Greek letters $ \rho, \tau, \sigma$ and the partial operator $\partial$).

Since this is a {\em one-to-one map}, it is clear that all relations written in one form can 
also be transcribed into the other, and in principle no form can be given preference.\\
However, the crucial difference is, that one has to put a postulated {\em metric tensor} on top of
the vector space, to define a {\em vector norm} and get covariant equations there (this is 
the definition of a {\em tensor space}).\\ 
As shown below, for the matrix representation the existence and form of this tensor 
is a consequence of the algebraic structure.

For binary matrices holds:\FN{
  the ``bar'' operation $\tau \to \bar \tau$ stands for matrix adjunction
  and $|\tau|$ for the determinant of the matrix $\tau$, i.e. $|\tau| \tau^{-1} = \bar\tau$ holds.
  $\TR\tau $ here denotes the scalar trace of $\tau$, and from
  $\tau\bar \tau = \bar\tau \tau = |\tau|I$ follows $|\tau| = \frac 12\TR{\tau\bar\tau}$.
}  
$\bar \MV x = x^\mu \bar \tau_\mu$ and consequently
the {\em matrix determinant} naturally defines a quadratic norm in $(x^\mu)$. This
norm can now be identified with the  norm of the vector space.
This is  only for 2x2 matrices possible, and vector dimensions greater than four are excluded.\\

The metric tensor $g = (g_{\mu\nu})$ is then given by:
\begin{equation}
  \label{eq-metric}
  |\MV x| = 
  x^\mu x^\nu \underbrace{\frac 12\TR{\tau_\mu\bar \tau_\nu}}_{\DEF g_{\mu\nu}}  = 
  x^\mu x^\nu g_{\mu\nu}.
\end{equation}

Obviously, symmetry follows $g_{\mu\nu} =g_{\nu\mu}$ and all are real numbers,
as required.

On the other hand, the four matrices $\tau_\mu$  (like every hermitean matrix) can be expressed as 
linear combinations of the 3 Pauli-matrices $\sigma_1,\sigma_2, \sigma_3$ and 
a fourth matrix $ \sigma_0 \DEF I = {10\choose 01}$:
\begin{equation}
  \label{eq-tau-sigma}
  \tau_\mu = a_\mu^\nu \sigma_\nu,
\end{equation}
with 16 real coefficients $a_\mu^\nu$ (for the vector space components this is to regard as a 
coordinate transformation: $x^\mu \to a^\mu_\nu x^\nu$).\\
Then follows from the known characteristics of the Pauli-matrices:\FN{
  with the usual representation
  $\sigma_1 = {0,1\choose 1,0}, \sigma_2 = {0,-i\choose i,\;\;\; 0}, \sigma_3 = {1,\;\;\; 0\choose 0,-1}$
  one easily checks  for all pairs $\mu,\nu = 0,\dots,3$ the orthogonality relation: 
  $\sigma_\mu \bar\sigma_\nu + \sigma_\nu \bar\sigma_\mu = I g_{\mu\nu}^{(0)}$.
}

\begin{equation}
  \label{eq-metr-sigma}
  g_{\mu\nu} = a_{\mu}^0 a_\nu^0 - a_{\mu}^1 a_\nu^1 - a_{\mu}^2 a_\nu^2 - a_{\mu}^3 a_\nu^3 =
  a_{\mu}^\lambda a_\nu^\delta g_{\lambda\delta}^{(0)},
\end{equation}
with $ g^{(0)} = (g_{\lambda\delta}^{(0)}) = diag[+1,-1,-1,-1]$ 
as conventional {\em \SC{Minkowski} metric tensor}.
From this equation follows, that all possible metric tensors are transformations of $g^{(0)}$ and locally
this metric can always be chosen. 
If the restriction of metric invariance ($g = g^{(0)}$) is made, then the $(a_\mu^\nu)$ 
are identical to the {\em \SC{Lorentz}-group}.\\

Consequently for simplification, the set of Pauli-matrices $\sigma_\mu$ is used in the following as basis.
In this case the components can be simply recovered from the matrix form $\MV x = x^\mu\sigma_\mu$ by

\begin{equation}
  \label{eq-comp-recover}
  x^\mu = \frac 12 \TR{\MV x \sigma_\mu} \qquad \Longleftrightarrow\qquad \MV x = x^\mu\sigma_\mu.
\end{equation}
Explicitly it has the simple form  $\MV x =\displaystyle {\; t+z,\;\;\; x-iy \choose x+iy,\;\;\; t-z}$.\\
  
Because the matrix algebra includes addition and subtraction operations, also trivially the symmetry under
{\em spacetime translations} holds, i.e. it shows the complete \SC{Poincare} group symmetry.

\section {Transformations and Covariant Forms}
A {\em \SC{Lorentz}-transformation} is represented here by an unimodular $2\times 2$ matrix $T \in SL(2,\mathcal C)$, 
$|T| = 1$ and
a \SC{Minkowski}-matrix transforms with:\FN{
  $\HC T$ denoting the conjugate transpose (or hermite conjugate) of $T$.
}
\begin{equation}
  \label{eq-trafo}
  \MV x \to \MV x' = T\MV x \HC T,
\end{equation}
which obviously preserves the hermitecity and the \SC{Minkowski}-invariant $|\MV x|$.
It has of course 6 free real (3 complex) parameters.\FN{
  This group is {\em homomorphic} to the restricted \SC{Lorentz}-group 
  and the homomorphism possesses the {\em kernel} $T \in \{ I,-I \}$ (see e.g. \cite{PENROSE}, pp. 16).
}

The general product $\MV A \bar \MV B$  (obviously $\MV A \MV B$ is not covariant under proper LT) 
of any two \SC{Minkowski}-matrices $\MV A, \MV B$ is then apparently 
a covariant matrix, because it transforms with:
\begin{equation}
  \label{eq-mmprod}
  \MV A \bar \MV B \to (T \MV A \HC T) (\bar \HC T \bar \MV B \bar T) =  T (\MV A \bar \MV B) \bar T.  
\end{equation}
The general scalar-product is the invariant expression, which is evidently always real:\FN{
  Above product matrix (\ref{eq-mmprod}), can be decomposed into two covariant expressions:
  a scalar commutator (which is this scalar-product) 
  and a traceless anti-commutator 
  $\MV A \bar \MV B = \frac 12 (\MV A \bar \MV B + \MV B \bar \MV A) + \frac 12 (\MV A \bar \MV B - \MV B \bar \MV A)$.
}
\begin{equation}
  \label{eq-scalarpr}
  \frac 12 \TR{\MV A \bar \MV B} = A_\mu B^\mu.
\end{equation}

Space rotations, as important special case, are the subgroup of matrices, obeying 
$\HC T = \bar T (\equiv T^{-1}) $.
They also preserve the trace, which represents the time component $x^0 = \frac 12\TR {\MV x}$.\FN{
  Since $T$ is then a similarity transformation, $T \MV x T^{-1}$, it is also clear that both eigenvalues of
  $\MV x$ are invariant.
}\\

Another important tranfsormation, which cannot be represented with any matrix $T$ of this group, 
is {\em spatial inversion} $\mathcal P$. It is obviously described by\FN{
  since $\bar \sigma_0 = \sigma_0$ and $\bar \sigma_1 = -\sigma_1, \dots$
}
\begin{equation}
  \label{eq-space-invers}
  \MV x \to \MV x_{sp} = \bar \MV x.
\end{equation}

It is remarkable, that  $\mathcal P$ is closely connected to the matrix multiplication order,
since a general covariant equation of the form $\MV A \bar \MV B = C$ transforms to $\bar \MV B \MV A = \bar C_{sp}$.

\section{Relativistic Electromagnetics}
\subsection{\SC{Maxwell}-Equations}
At first I will shortly list \SC{Maxwell}s equations  in {\bf component notation}.\\
Contemporary textbooks usually start the derivation of relativistic electrodynamics with the 
{\em 4-vector potential} $A^\mu$, where $A^0 = V$ is the electric and $\vec A = (A^1,A^2,A^3)$ the 
magnetic potential.

The antisymmetric {\em field strength} tensor $F_{\mu\nu}$ is then derived 
from $A_\mu$ with the ansatz $F_{\mu\nu} = \PD [A_\mu]{x^\nu} - \PD [A_\nu]{x^\mu}$. 
It is composed from electric and magnetic field vectors $\vec E, \vec B$:
\[ F_{01} = E_1,\dots \MBOX{and}  F_{12} = B_3,\dots,.
\]
Then the 4 {\em inhomogeneous} \SC{Maxwell}-eqs. are (with $J_\mu$ as 4-vector  of current, 
see e.g. \cite{EINSTEIN}, p. 42)
\[ \PD [F_{\mu\nu}]{x_\nu}  = J_\mu.
\] 
The 4 {\em homogeneous} eqs. however, can be derived from the above potential ansatz for $F$

\[ \PD [F_{\mu\nu}]{x_\sigma} + \PD [F_{\nu\sigma}]{x_\mu} + \PD [F_{\sigma\nu}]{x_\nu}= 0.
\]

In {\bf matrix form}  the {\em vector potential} is obviously represented by the \SC{Minkowski}-matrix 
$\MV A = A^\mu\sigma_\mu$, using the general mapping formula (\ref{eq-comp-recover}).

The {\em field strength matrix} $F$ is here derived from a general covariant product, similar to 
(\ref{eq-mmprod}),
with the {\em partial derivation operator} $\partial$ in the form $\bar \partial \MV A$. 
$\partial$ is according to (\ref{eq-comp-recover}) a hermitean (\SC{Minkowski}-) matrix with the explicit form
\begin{equation}
  \label{eq-partial}
  \partial \DEF \sigma_\mu \partial^\mu =  \sigma_\mu \PD {x_\mu} =
  \PD {x_0} + \sigma_1 \PD{x_1}+ \cdots =  \PD {t} + \nabla.
\end{equation}

We use the anticommutator of this form to get a traceless matrix $F$,  ($F + \bar F = 0$) :
\[
 F \DEF \frac 12(\bar\partial \MV A - \bar\MV A \partial).
\]
It is then easy to show, that $F$ (it has 3 complex = 6 real components) combines the field vectors, 
here both as traceless, hermitean matrices 
$E = E^k\sigma_k, (k=1,\dots, 3)$ and $B = B^k\sigma_k$:
\begin{equation}
  \label{eq-F-EB}
  F = E + i B.
\end{equation}
Now \SC{Maxwell}s equations are represented by only \underline{\em one matrix equation}, which 
includes either homog. and inhomog. 
eqs.:\FN{
  This matrix eq. consists of 4 complex, i.e. 8 real eqs. In chapter \ref{sec-lagrangian} is shortly 
  sketched, how it can be derived from a \SC{Lagrangian}.
}
\begin{equation}
  \label{eq-maxwell-matrix}
  \underline {\partial F = \MV J}.
\end{equation}
\underline{Proof}:
The l.h.s. of eq. (\ref{eq-maxwell-matrix}) can be decomposed into an hermitean and 
anti-hermitean term (vanishing, since $\MV J$ is hermitean),
which are {\em both} \SC{Maxwell} eqs. 
\[
\partial F = (\PD t +\nabla)(E+iB) = 
\underbrace{\dot E + \nabla\cdot E + i\nabla \times B}_{\stackrel != \MV J} + 
\underbrace{\nabla \times E + i \dot B + i \nabla\cdot B}_{\stackrel != 0}\MBOX{\underline{q.e.d.}}
\]
The {\em \SC{Lorentz}-covariance} of  (\ref{eq-maxwell-matrix}) is guaranteed, when the following 
transformation rule for $F$ is 
assumed\FN{
  Like necessary, for space rotations $\bar T = \HC T$ then $E,B$ transform independently as 3-vectors,
  but for proper LT, they get mixed.
}
\begin{equation}
  \label{eq-trafo-F}
  F \to F' = \bar \HC T F \HC T.
\end{equation}
For checking the mirror-invariance of  (\ref{eq-maxwell-matrix}) one must realize, that $E,B$ transform as
proper- and pseudo-vectors, resp. under spatial inversion: $E_{sp} = \bar E = -E$ and   
$B_{sp} = -\bar B = +B $. Thus $F_{sp} = -\HC F = \bar \HC F$ holds and consequently
 (\ref{eq-maxwell-matrix}) is mirror-invariant.\FN{
   From $\partial_{sp} F_{sp} = \MV J_{sp}$ $\to$ $\bar\partial \bar \HC F = \bar \MV J$, and after
   bar-operation and herm. conj. one gets the original eq. again, q.e.d.
}\\ 

{\bf Discussion:} The main difference between matrix- and component form is, that the homogeneous eqs.
are not needed as independent assumptions. They are fulfilled in any case, regardless of the 
potential ansatz.

\subsection{\SC{Lorentz}-Force}
In conventional component form the \SC{Lorentz}-force is $K_\mu = F_{\mu\nu}J^\nu$. 
Here one has the matrix form, which obviously gives a hermitean force matrix $\MV K$:
\begin{equation}
  \label{eq-lor-force}
  \MV K = \frac 12 (\MV J F + \HC F \MV J).
\end{equation}
Of course, it is \SC{Lorentz}-covariant and mirror-invariant.

\subsection{Energy-Momentum-Tensor of Electromagnetic Field}
Although it is not strictly necessary for the main thesis of this paper, I included this
chapter, because it shows quite impressively the power of the matrix formalism.\FN{
  Of course, a general tensor with 16 real components, or a symmetric tensor with 10, cannot be represented 
  by a single 2x2-matrix, but only by a set of matrices.
}\\

Inserting the \SC{Maxwell} eq. (\ref{eq-maxwell-matrix}) into the \SC{Lorentz}-force (\ref{eq-lor-force}) 
immediately gives:\FN{
  the parentheses in the first terms denote the differential-operands of $\partial$, while in the underlined term it
  operates both to the left and right 
}
\begin{equation}
  \label{eq-EM-force}
 \underline{\MV  K} = \frac 12 ((\HC F\partial) F + \HC F (\partial F))= \frac 12 \underline{\HC F \partial F} = 
  \PD{x_\mu} \underbrace{\frac 12(\HC F\sigma_\mu F)}_{\DEF \MV T_\mu} =  \underline{\PD [\MV T_\mu] {x_\mu}}.
\end{equation}
This derivation, consisting only of two simple reorderings, is significantly more concise than the 
corresponding component form (see e.g. \cite{EINSTEIN}, p. 50).
Obviously the four hermitean matrices $\MV T_\mu$ (with 16 real components) here represent the 
{\bf energy-momentum tensor}.\\

To get the corresponding component form, one uses the general mapping formula (\ref{eq-comp-recover}), 
which here leads to the 16 real components:
$T^\nu_\mu = \frac 12 \TR{\MV T_\mu\sigma_\nu} $.\FN{ and $\MV T_\mu = T^\nu_\mu\sigma_\nu$ }\\
Then with the following explicit formula the symmetry of $T_{\mu\nu} =T_{\nu\mu}$ can be easily shown,
with usual formulas for the trace:
\begin{equation}
  \label{eq-EM-expl}
  T_{\mu\nu} = \frac 12 \TR{\MV T_\mu\bar \sigma_\nu}  = \frac 14 \TR{\HC F\sigma_\mu F\bar \sigma_\nu}.
\end{equation}

\section{Relativistic Quantum Mechanics}

In this section I will show, that relativistic quantum mechanics 
can be readily expressed with $2\times 2$ matrices (which is well-known for a great part), 
but useless degrees of freedom are significantly supressed.\\ 

This is an especially important case, since in our contemporary understanding, quantum mechanics
and esp. \SC{Dirac}s eq. (with its various generalizations) is the fundament of the physical world.
 On the other hand, this theory is surely
not yet finished, and it is to expect that new insights will evolve in the future,
possibly within the framework of the matrix formalism.\\

Here closes the circle: the matrix formulation was first introduced for the 
description of quantum mechanical spin and can now hopefully lead to a better understanding of physics.\\

\subsection{\SC{Dirac}-Equation}
\label{sec-dirac}
In most modern textbooks \SC{Dirac}s eq. is presented in the conventional {\bf component notation}, with the 
four \SC{Clifford} matrices $\gamma_\mu$ 
(and $\partial^\mu \DEF \PD {x_\mu}$) for the 4-spinor wave function as column vector 
$\psi = (\psi_1,\dots,\psi_4)^T $: (see \cite{FEYNMAN}, p. 50, \cite{LL} pp. 110)
\begin{equation}
  \label{eq-dirac-comp}
  i\gamma_\mu \partial^\mu \psi = m\psi.
\end{equation}
This is a mathematically very elegant form, but it is achieved at the price of 
loss of physical reality of $\psi$. 
It exposes a great amount of ambiguity, since it is obviuously invariant under the so called
{\em similarity transformations}\FN{ 
  They are not connected to a \SC{Lorentz}-transformation, since the spacetime 
  components are not affected at all.
}(see e.g. \cite{FEYNMAN}, p. 55):
\[
   \gamma_\mu \to U\gamma_\mu U^{-1} \MBOX{and} \psi \to U\psi. 
\]
Here $U$ is an arbitrary $4\times 4$ matrix,  containing 16 free complex parameters.  
This means, the formula (\ref{eq-dirac-comp}) allows a linear transformation, leading to different
representations, with 16 complex parameters without
any change of the physical meaning. The components of $\psi$ thus cannot represent any 
physical entities directly.
In my opinion, this is a great disadvantage of this formula.\\

For the derivation of the {\bf matrix form}, I start with the {\em \SC{Weyl}-representation} of the $\gamma_\mu$
\begin{equation}
  \label{eq-Weyl-gamma}
  \gamma_0 = {0, -I_2\choose -I_2,0},\MBOX{and}
  \gamma_k = {\;\;\; 0,\;\; \sigma_k\choose -\sigma_k, 0},\quad k=1,2,3.
\end{equation}
This form has the important special feature, that here the 4-spinor can be decomposed into two 2-spinors
$\Psi,\Phi$:
$\psi = {\Psi \choose \Phi}$, which transform independently under \SC{Lorentz}-transformations (see below),
 and (\ref{eq-dirac-comp}) reads with them:
\begin{equation}
  \label{eq-zig-zag} 
  i\partial\Phi = -m\Psi\MBOX{and} i\bar \partial \Psi = -m\Phi.
\end{equation}
An additional, external electromagnetic vector potential field $\MV A$ is as usual introduced by the 
substitution $\partial \to \partial - ie \MV A$:
\begin{equation}
  \label{eq-zig-zag-A} 
  (i\partial + e\MV A)\Phi= -m\Psi\MBOX{and} (i\bar \partial +e\bar \MV A) \Psi = -m\Phi.
\end{equation}
This bi-spinor form of \SC{Dirac}s eq. is well-known (although in most cases given in 
slightly different notation, see e.g. \cite{LL}, p. 70) 
and sometimes referred to as "zigzag" model of the electron (e.g. \cite{PENROSE}).
From (\ref{eq-zig-zag-A}) the \SC{Lorentz}-transformation rules for the 2-spinors can be derived as:\FN{
  consider again that $\partial$ and $\MV A$ transform like $\partial \to T\partial \HC T$
}
\begin{equation}
  \label{eq-trafo-2-spin}
  \Psi \to T\Psi \MBOX{and} \Phi  \to \bar \HC T\Phi,
\end{equation}
leading to obviously covariant eqs. (\ref{eq-zig-zag-A}).
Under spatial inversions both eqs. and consequently the spinors are
interchanging:  $\Psi \leftrightarrow \Phi$.\\

\subsection{Matrix-\SC{Dirac}-Equation} 
It is not yet commonly known, however, that both parts of (\ref{eq-zig-zag-A}) can be combined in
one {\em single matrix equation}. This representation must be considered as the natural form
of \SC{Dirac}s eq. in the MST context, and it opens up new possibilities for its generalization.\\

To develop this matrix eq., the second equation of (\ref{eq-zig-zag-A}) is converted in the following manner.
With $M \DEF i\partial + e\MV A$ it reads $\bar M\Psi = -m\Phi$.\\
Now one uses the general identity for every 2x2 matrix $M$ ($M^T$ denoting transposed matrix):\FN{
  A geometric explanation is, that the bar-operation as mentioned already, means
  spatial inversion, which is equal to the combined operation
  of transposing (i.e. $y \to -y$) and a rotation around $y$ of 180°, which is performed 
  by the transformation $T  = \rho = i\sigma_2 $. 
}
\[ 
\bar M = \rho M^T \bar \rho, \MBOX{with}\rho \stackrel {def} ={\;\;0,\;1\choose -1,0}
\]
The auxiliary matrix $\rho$ is sometimes denoted as "spinor metric", because it defines 
an invariant spinor determinant (see chapter \ref{2-sp-and-mm}).
Inserting the above identity leads to $ M^T\rho \Psi = -m \rho \Phi$.

Of this one takes the complex conjugate, using $(M^T)^\ast = \HC M = -i\partial + e\MV A$:
\[
  (-i\partial + e\MV A) \rho \Psi^\ast = -m\rho \Phi^\ast.
\]
Here it is obviuously useful to define a new "tilde-operator"\FN{ 
  this operator obeys $\HCB {\HCB \Psi} = -\Psi$, since $\rho^2 = -1$
} 
for 2-spinors: $\underline{\HCB \Psi \DEF \rho \Psi^\ast}$ and the last eq. then writes
$ (-i\partial + e\MV A)\HCB \Psi = -m\HCB \Phi$.\\ 
Then it is possible to combine this equation and the first of (\ref{eq-zig-zag-A}) as 2 columns into one 2x2 matrix equation:
\[
  e\MV A(\Phi,\HCB\Psi) + i\partial(\Phi,-\HCB\Psi) = -m(\Psi,\HCB\Phi).
\]
Now one defines the ``spinor-matrix'' $ P \stackrel {def}= (\Phi,\HCB\Psi)$ 
(which is the replacement of the 4-spinor $\psi$) and 
notes $\bar{\HC P} = -(\Psi,\HCB\Phi)$, and with the constant matrix 
$S \stackrel {def}={i,\;\;0\choose 0,-i} $ finally gets:
\begin{equation}
  \label{eq-dirac-matrix} 
  \underline { e\MV A P + \partial PS = m\bar{\HC P}}.
\end{equation}
Although this formula at a first glance looks somewhat uncommon, 
esp. the right-side factor $S$ in the derivation term, it possesses all features and solutions 
of the original 4-spinor equation (\ref{eq-dirac-comp}).\\
The 2x2-matrix $S$ together with the operator on the r.h.s here "magically absorb" 
all 4 \SC{Clifford} matrices $\gamma_\mu$. It should be clear from the above, that the special 
form of $S$ ($S = i\sigma_3$), is the consequence of the choice of $\gamma_\mu$. A more general form shall
be discussed below.
\\

To demonstrate the novel power of this matrix eq., one can derive an equivalent bilinear form by multiplicating 
it from left\FN{
  multiplication from right produces another eq. with the same r.h.s. 
}
with $\HC P$, resulting in 
\BE e\HC P \MV A P + \HC P (\partial P) S = m \CC{|P|}.
\EE
Note, that this is still a matrix eq., although the r.h.s. is scalar ($\sim I$) 
and the l.h.s. terms are \SC{Lorentz}-invariants. And it is still {\em equivalent} to (\ref{eq-dirac-matrix}), 
provided $P$ is not singular ($|P| \ne 0$).\\
This direct way is
only possible by using matrix algebra. By utilizing this bilinear form, esp. many computations, 
e.g. regarding gauge invariance, \SC{Lagrangian} and conservations laws can be performed much simpler.\\

According to above definitions, $P$ transforms consistently with 
${P\to \bar \HC{ T} P}$ under \SC{Lorentz}-transformations and ($\ref{eq-dirac-matrix}$) 
is obviously covariant. 
Since $\bar \HC T $ operates only from the left on $P$, the two column 2-spinors of $P$ 
transform equally and independently.\\
The {\em mirror-invariance} is guaranteed with $P \to P_{sp} = \bar\HC P$ (since $S = \bar \HC S$).\\

Here also a {\em similarity transformation} is possible by right-side multiplication\FN{
  left-side multiplication always describes a \SC{Lorentz}-transformation
} 
of $P$ with a matrix $U$ obeying $U = \bar\HC U$
\begin{equation}
  \label{eq-matr-sim}
  P \to PU \MBOX{and} S \to  U^{-1} S U,
\end{equation}
but this 2x2-matrix $U$ has only 2 free complex parameters (4 real), compared to 16 above 
(since one of the 4 real parameters is only a constant factor, there actually remain only 3 real free 
parameters).\\
Essentially this transformation says, that $S$ (like $U$) can be any matrix obeying the condition $S = \bar \HC S$, 
which describes a subalgebra of matrices, which is isomorphic to the algebra of {\em quaternions}.

An obvious possibility to explain the remaining ambiguity physically, is discussed 
in chapter \ref{sec-yang-mills}.\\

The {\em gauge invariance} of (\ref{eq-dirac-matrix}) and the corresponding \SC{Lagrangian} 
(\ref{eq-lagr-dem}) below is a  bit different to the conventional form, because $P$ {\em cannot} be
multiplied with a scalar complex phase factor $e^{i\lambda}$, because the mass-term would then
transform with  $e^{-i\lambda}$. This impossibility to apply scalar phase factors is probably the reason, 
that this quite simple and obvious formula has never been considered before.  
Also the usual covariant replacement of the derivation operator 
$\partial^\mu \to D^\mu = \partial^\mu - ie A^\mu$ cannot simply be transcribed
to $D = \partial - i\MV A$, but must be modified here.\\
However, one easily checks, that e.g. the gauge transformation
\BE
  \label{eq-gauge}
  P \to Pe^{\lambda S} \MBOX{and} e\MV A \to e\MV A + \partial \lambda,
\EE
where $\lambda(\MV x)$ is an arbitrary real spacetime function, is the correct form.\FN{
Note that  $e^{\lambda S}$ commutes with $S$ and $\partial e^{\lambda S} = (\partial\lambda) e^{\lambda S} S$.
}\\
The matrix $S$ can be thus seen as replacement of the imaginary unit $i$, since it also obeys $S^2 = -1$.\\

Stationary states, which represent bound states in atoms, are here similarly 
described by the ansatz $P = P_0(\MV r) e^{-\varepsilon t S}$ (with $\varepsilon$ as energy), 
which  results in (since $\PD {t} P = -\varepsilon P S$)
\[(\varepsilon + e\MV A )P_0 +\nabla P_0 S = m \bar\HC P_0,
\]
and it is easy to show, that it has the the
same solutions as the original \SC{Dirac}-eq.\\
 
An important special case regards massless, uncharged fermions, i.e. {\em neutrinos}. 
It is known from experiments, that only {\em left-handed} neutrinos exist, 
right-handed ones have never been observed.
 
\SC{Dirac}s original eq., eg. written in the form (\ref{eq-zig-zag}), leaves 
this fact unexplained, because both parts decouple with $ m = 0 $, and so they have independent 
solutions for $\Psi$ and $\Phi$, representing both types of neutrinos.

The non-existence of right-handed neutrinos is a 
{\em direct consequence of the matrix eq.} (\ref{eq-dirac-matrix}), however.\\
With $m = 0, e= 0$ it simplifies to $\partial P S = 0$ 
(for this singular case, (\ref{eq-dirac-matrix}) is no longer {\em equivalent} to the original
\SC{Dirac}-eq.).
Here the factor $S$ can be eliminated (by rhs multiplication with $S^{-1}$) giving 
$\underline{\partial P = 0}$. All their solutions have left-handed chirality, which is easy to show
by transforming it into momentum space.\\

In close connection to this, also {\em weak interactions} in the V-A-theory are most simply expressed in this 
form (\ref{eq-dirac-matrix}). 
It follows from the fact, that for the 
used \SC{Weyl}-representation the matrix $\gamma_5 \DEF i\gamma_0\gamma_1\gamma_2\gamma_3$ is a diagonal matrix:
$\gamma_5 = {I_2,\;\;\;0 \choose 0,-I_2}$. 
And since weak interaction couples in the 4-spinor form 
with $I_4 \pm \gamma_5$, so always in one of the eq-pair (\ref{eq-zig-zag-A}) 
the respective term vanishes.\\
Further considerations, regarding  electro-weak gauge theory are done in chapter \ref{sec-yang-mills}.

\subsection{\SC{Lagrangian} of Coupled \SC{Dirac}- and EM-Field}
\label{sec-lagrangian}
\SC{Lagrangian}s play a very important role in modern
field theory.
They can readily be written in matrix form using the above entities. 
For the combined \SC{Dirac}- and em-field it is the sum of four scalar terms:
\BE
\label{eq-lagr-dem}
\LAGR = \TR{\HC P(\partial P) S} + e\TR{\MV A P\HC P } - 2\Re|F| - 2m \Re|P|.
\EE
This form demonstrates another advantage of the matrix representation. 
It can reveal subtle similarities between some terms (here e.g. the 3. and 4. term), which are hidden in the
component form.\\
The validity of (\ref{eq-lagr-dem}) can be proved by transforming it into component form, or better by deriving the field
eqs., namely (\ref{eq-maxwell-matrix}) and (\ref{eq-dirac-matrix}) from it. 
This complete derivation must be omitted here, only some basic steps should be stated.

In the first (differential) term, the partial operator should only operate to the right (as indicated
by the parentheses). 
Furthermore one notices, that this term is not real (as 
normally required for a \SC{Lagrangian} and is the case for the other three terms).
However, the actually relevant spacetime integral {\em is real}:
\[ \mathcal I = \int d^4 x\; \TR{\HC P(\partial P) S} = \mbox{real,} \MBOX {i.e.} \Im \mathcal (I) = 0,
\] which is proved with the vanishing of the integral 
$ \int \TR{(\HC P\partial P) S} = \int \partial_\mu \TR{\HC P\sigma_\mu P S} = 0$ by 
{\em \SC{Gauss}' law} and partial integration.\\

The third term  is the well-known \SC{Lagrangian} 
of the electromagnetic field $\LAGR_{em} = -2\Re|F| = E^2 - B^2$, since 
$F =\frac 12(\bar\partial \MV A - \bar \MV A\partial) = E+iB$.\\

Consequently, the variation of $\MV A$ in the second and third term, leads to \SC{Maxwell}s eq. 
(\ref{eq-maxwell-matrix}), if 
the 4-current of the \SC{Dirac}-field is defined as 
\BE \label{eq-P-current} 
\MV J_e \DEF e \overline{P\HC P} \quad (= e\bar\HC P \bar P).
\EE
Variation of $\HC P$ (or independently $P$) in the terms 1, 2 and 4  leads to \SC{Dirac}s-eq. (\ref{eq-dirac-matrix}).

\subsection{\SC{Yang-Mills} Gauge-Fields in Matrix Form}
\label{sec-yang-mills}
The remaining possibility of similarity transformations $P \to P U$ ($\bar U = \HC U$, see chapter \ref{sec-dirac}) 
leads to an obvious generalization of the \SC{Dirac} eq. in matrix form
(\ref{eq-dirac-matrix}) with four vector fields $\MV B^\mu,\; \mu = 0,\dots,3$:
\BE
\label{eq-yang-mills}
\partial P S + \MV B^\mu P \sigma_\mu = m \bar\HC P.
\EE
This also resolves the remaining ambiguities of $P$.
Here $\MV B^0 \equiv \MV A$ is apparently again the em. vector potential, which is invariant 
under this transformation.
The other three fields $(\MV B^1,\MV B^2,\MV B^3)$ mix however (under \SC{Lorentz}-trafos they
act as normal MMs, like $\MV A$), since with $U\sigma_k U^{-1} = a^m_k \sigma_m,\;\; (m,k = 1,2,3)$ follows
$\MV B^k \to a^k_m \MV B^m$. With the additional restriction $|U| = 1$ this is the $SU(2)$ group and its
 $SO(3)$ representation acts on the $\MV B^k$.\\

 One can now formulate the interesting hypothesis, that by introducing a local non-abelian gauge field
$U(\MV x)$ the unified electro-weak field may be represented, similar to \SC{Yang-Mills} theory 
(see e.g. \cite{EBERT}).\\

It is striking, that this gauge field shows remarkable similarities to the symmetry $SU(2)\times U(1)$ as 
proposed by \SC{Weinberg} and \SC{Salam} for the unified theory, although it is evidently
not equivalent.\FN{
  Here  $U(1)$ for electromagnetic gauge is a subgroup of $SU(2)$.
}\\  
To get all 4 gauge fields, it is obviously necessary to use the infinitesimal
generator of the complete quaternionic algebra\FN{
  See again chapter \ref{sec-dirac}; 
  the only required condition for $U$ is actually $\bar U = \HC U$, which is fulfilled by this algebra.\\
  Remember, that $\sigma_0 = I$, so one gets 
  $\bar U = \HC U  = e^{\lambda^0 -i\lambda^k\sigma_k } $
} for $U$ instead of the subset $SU(2)$, which is:
\[
   U(\MV x) = e^{\lambda^0 \sigma_0 +i\lambda^k\sigma_k } \approx I + \lambda^0\sigma_0 + i\lambda^k\sigma_k,\; k=1,2,3, 
\] 
with 4 real spacetime functions $\lambda^\mu(\MV x),\;\; (|\lambda^\mu| \ll 1)$.\\
Then $\lambda^3(\MV x)$ represents the em. gauge field, coupled with $\MV A \equiv \MV B^0$ (if $S = i\sigma_3$,
as before (\ref{eq-gauge}) explained).\\
The gauge fields $\lambda^{1,2}(\MV x)$ couple with $\pm \MV B^{2,1}$, respectively.
The gauge field $\lambda^0(\MV x)$ obviously represents a boost (since $|U|\ne 1$) and couples with $\MV B^3$, 
very different to the standard theory.

In conclusion should be emphasized the remarkable fact, 
that the gauge symmetry here is an {\em intrinsic} feature of \SC{Dirac}s eq. in matrix form
and its group structure is automatically determined. Moreover, gauge- and \SC{Lorentz}-symmetry here turn 
out to be
``two sides of one coin'' in the general transformation formula for the spinor-matrix $P \to T P U$. 

Further discussions, regarding covariant field equations for the associated generalized em. field tensors 
$F^\mu$, the complete \SC{Lagrangian} and a possible \SC{Higgs} mechanism for symmetry breaking,
go beyond the scope of this article and shall be considered in a subsequent paper.
 
\subsection{2-Spinors and \SC{Minkowski}-Matrices}
\label{2-sp-and-mm}
At the end, some general remarks about the relations of spinors and matrices should be added.
As stated above, 2-spinors are represented by binary column matrices 
$\Psi = {\alpha\choose\beta}, \Phi = {\gamma\choose\delta}, \dots$, which
transform under LT as $\Psi \to T\Psi$. 
Then for a any pair of spinors $P = (\Psi,\Phi)$ the determinant
\[ 
|P| = |\Psi,\Phi| = \Psi^T\rho\Phi = \alpha\delta -\beta\gamma 
\] 
is obviously a \SC{Lorentz}-invariant, because
$TP = (T \Psi,T \Phi)$.
 
Also note the important fact, that spinor products, like e.g. the matrix:\FN{ 
  In most textbooks a ``dotted index'' notation is used to describe conjugated spinors like 
  $\HC \Psi$, that
  goes back to the first publications on this topic.
  I do not adopt it here.
} 
\BE
\label{eq-MM-spinor}
\MV H = \Psi \HC\Psi =  {\alpha\choose\beta}(\CC\alpha, \CC\beta) =  
    {|\alpha|^2, \alpha\CC\beta \choose \beta\CC\alpha, |\beta|^2}
\EE
is obviously a \SC{Minkowski}-matrix (in this special example a null-matrix: $|\MV H| = 0$).
That says, that matrices can be constructed by spinors, but the opposite does not hold.
Only null-matrices can be uniquely (up to a phase-factor) decomposed into spinors.

It is also a fascinating feature of forms like (\ref{eq-MM-spinor}), 
that they have a positive definite time-component, which might help to explain the direction of time.
From the realization, that 2-spinors are the algebraic basis of all, 
it will be possibly feasible to develop a complete theory of spacetime only with spinors.

The crucial problem is however, how to retain the 
spacetime translation symmetry in such constructs.

\section{Conclusions}
In this paper I have presented the most important concepts of Special Relativity in 2x2-matrix form, 
namely the entities and equations of electromagnetic interactions and the \SC{Dirac} equation.
Essentially this form uses another algebraic concept of spacetime, rather 
than the conventional vector space.\\

Although the equations are obviously equivalent to conventional component formulation, I have
showed that the matrix form has several striking advantages, which suggest that this form should
be considered as the primary description of the physical world.\\

The main advantages can be shortly summarized:
\begin{itemize}
\item The metric tensor needs not to be postulated and spacetime can have no more than four dimensions
\item The \SC{Maxwell} equations are represented by a single equation rather than two independent
\item The \SC{Dirac} spinor field in the novel \SC{Dirac} eq. in matrix form 
  has much less of degrees of freedom without any physical meaning
  and this form can explain the non-existence of right-handed neutrinos 
\item A new type of \SC{Yang-Mills} gauge fields arises from the generalization of this
  matrix \SC{Dirac}s eq., which possibly can describe electro-weak interactions
\end{itemize}

From a heuristic point of view, from a bunch of theories which describe the same phenomena with
equal accuracy, the one with the least prerequisites should be given preference.\\

Another major intention of writing this paper was, to encourage other theoretical physicists, to find 
extensions of this concept for new theories.
Also I hope to be able, to present a new concept for quantum mechanics on the basis of this algebra,
which can replace the wave function by a discrete model. Some first ideas can be found 
in \cite{WOLFK}.

\section{Acknowledgments}
I want to thank Dr. Charles Francis and Dr. Peter Enders for their interest and helpful tips in
preparing this paper.

\end {document}